\documentclass[aps,floats,floatfix,amssymb,amsmath,prb,nofootinbib,twocolumn,superscriptaddress]{revtex4-2}

\usepackage{calc}
\usepackage{psfrag}
\usepackage{graphicx}
\usepackage{color}
\usepackage[dvipsnames]{xcolor}
\usepackage[unicode=true,pdfusetitle,
 bookmarks=true,bookmarksnumbered=false,bookmarksopen=false,
 breaklinks=false,pdfborder={0 0 0},backref=false,colorlinks=true]
 {hyperref}
\usepackage{geometry}
\usepackage{changes}

\begin{document}

\title{Spin Seebeck coefficient and spin-thermal diffusion in the two-dimensional Hubbard model}
\author{Jernej Mravlje}
\affiliation{Jozef Stefan Institute, Jamova 39, Ljubljana, Slovenia}
\author{Martin Ulaga}
\affiliation{Jozef Stefan Institute, Jamova 39, Ljubljana, Slovenia}
\author{Jure Kokalj}
\affiliation{University of Ljubljana, Faculty of Civil and Geodetic
  Engineering,  Jamova 2, Ljubljana, Slovenia} 
\affiliation{Jozef Stefan Institute, Jamova 39, Ljubljana, Slovenia}

\begin{abstract}
  We investigate the spin Seebeck coefficient $S_s$ in the square
  lattice Hubbard model at high temperatures of relevance to cold-atom
  measurements. We solve the model with the finite-temperature Lanczos
  and with the dynamical mean-field theory methods and find they give
  similar results in the considered regime.  $S_s$ exceeds the atomic
  'Heikes' estimates and the Kelvin entropic estimates drastically.
  We analyze the behavior in terms of a mapping onto the problem of a
  doped attractive model and derive an approximate expression that
  allows relating the enhancement of $S_s$ to distinct scattering of
  the spin-majority and the spin-minority excitations.  Our analysis
  reveals the limitations of entropic interpretations of Seebeck
  coefficient even in the high-temperature regime. 
Large values of $S_s$ could be observed on optical lattices. We also calculated the full diffusion matrix. 
 We quantified the spin-thermal diffusion, that is, the extent of the mixing between the spin and the thermal diffusion and discuss the results in the context of recent measurements of the spin-diffusion constant in cold atoms.
\end{abstract}
\maketitle

\section{Introduction}

Cold-atom systems on optical lattices provide a novel lens on poorly
understood transport regimes of correlated
electrons.  They can realize the
Hubbard model—the standard model of correlated electrons that
interact with an onsite repulsion $U$ and move on the lattice (hopping
$t$)—without real world complications, such as lattice
vibrations and disorder. Hence one can directly and quantitatively compare
the outcome of the experiment to those of the numerical solutions of
the Hubbard model~\cite{brown19,nichols19}.  Such a cross-verification
turned very successful in the measurements of the charge
diffusivity~\cite{brown19}.  Besides providing an important mutual
benchmark of the methods it led to a quantification of the  vertex
corrections~\cite{vucicevic19}. Intriguingly, related measurements of
spin diffusivity revealed a disagreement between the numerical methods
and the experiment~\cite{nichols19}. An independent numerical
investigation~\cite{ulaga20} confirmed the results of the theory but
disagreed with the experiment.

This disagreement thus calls for a close
inspection of the underlying assumptions of the experimental
analysis. One of the assumptions is that the spin-thermoelectric
effect is unimportant. This holds strictly at a vanishing magnetization,
but in the actual experiment this condition was only approximately met
as some spin imbalance $m_z=(n_\uparrow-n_\downarrow)/2 $ 
is seen in the measurements: $|m_z|\lesssim 0.05$~\cite{nichols19}. 
The strength of the spin-thermoelectric effect is quantified by the spin-Seebeck coefficient $S_s$ given by the ratio of the magnetic field and thermal gradient at the condition of vanishing spin current, $S_s=  \nabla{B}/\nabla{T}|_{j_s=0}$.
$S_s$ is a quantity that is relevant for spintronics applications~\cite{zutic04,Uchida2008,Adachi2013,hirobe17} but has to our knowledge  not been discussed for the Hubbard model (surprisingly, as this is the paradigmatic model of correlated electrons). The intention of our work is to fill this gap, establish how large $S_s$ is in the high-temperature regime and use that knowledge discuss whether the measurements of spin-diffusion in cold atoms could be influenced by the spin-thermoelectric effects.

Some intuition concerning $S_s$ could be expected from considerations that relate the
 ordinary charge Seebeck coefficient $S_c$ to thermodynamic quantities, such as the
high-temperature Heikes limit $S_c^H = \mu/T$ or the Kelvin formula $S^K_c
= d \mu/dT $ that relate the Seebeck coefficient to the
temperature dependence of the chemical
potential~\cite{silk09,peterson10,deng13,kokalj15}.  Namely, it was demonstrated that often at high temperatures the Kelvin
formula describes the Seebeck coefficient well~\cite{deng13} (but some
exceptions to this were also noted~\cite{mravlje16}).

To apply this intuition to  $S_s$ one can 
use  a mapping that relates the spin transport in a magnetized
repulsive Hubbard model to the charge transport in a doped attractive Hubbard
model~\cite{emery76,laloux94,keller01,capone02,toschi2005,moreo07,Kuleeva2014,osolin15}.
This mapping proceeds via a particle-hole transformation on particles
of only one spin, e.g, $\downarrow$, with
$c_{i,\downarrow} \rightarrow (-1)^ic_{i,\downarrow}^\dagger$ and results
in an interchange of spin and charge degrees of freedom, explicitly
$n_\uparrow - n_\downarrow \rightarrow n_\uparrow + n_\downarrow$,
$n_\uparrow n_\downarrow \rightarrow - n_\uparrow n_\downarrow$. Accordingly, Hubbard
repulsion goes to attraction, $Un_\uparrow n_\downarrow \rightarrow -U n_\uparrow n_\downarrow$. 
Via this mapping, one can
relate the spin Heikes estimate for spin Seebeck coefficient  $S_s^H =
B/T |_{m_z=\mathrm{const}}$ (or Kelvin estimate $S_s^K = dB/dT |_{m_z=\mathrm{const}}) $ to the corresponding charge Heikes and Kelvin estimates for a model with an opposite
sign of repulsion (that is, an attractive model for the case of interest
here).  Actually, by exploiting the mapping one can use the results from
the literature~\cite{chaikin76} and obtain $S_s=8k_B m_z $ at a high
temperature ($T>U$) and $S_s=4k_B m_z$ at a lower temperature
($T<U$). From these estimates—that one is inclined to trust,
especially in the high-temperature regime $T>t$ pertinent to cold atom
measurements—one expects only small values of the spin Seebeck
coefficient $S_s \ll k_B$, since $m_z \ll 1$.

 In this paper we show that this reasoning is incorrect.  We calculate
$S_s$ for a square lattice Hubbard model at
high-$T$ using the  finite temperature Lanczos  method (FTLM) and the dynamical mean-field
theory (DMFT) approaches and find that it strongly exceeds the bounds just
discussed. $S_s$ violates the thermodynamic expectations even in the high-temperature regime, an unexpected finding based on what was previously known for the ordinary thermoelectric effect. 
Large values of $S_s$ call for a reexamination of the possible
spin-thermoelectric effects in cold atom measurements of
spin-diffusion. To estimate those, we calculated the full diffusion matrix $\mathbf{D}$. 
In general, the eigenvalues of $\mathbf{D}$ deviate from those found in the absence of spin-thermoelectric effect. Interestingly, we find the deviations are related to the difference between the actual value of spin-Seebeck coefficient and its thermodynamic Kelvin approximate,  $S_s- S_s^K$.
We discuss why in spite of  this difference being sizable (it exceeds $k_B$ at large $U$) the final influence on the measured spin-diffusion for moderate magnetization  is unimportant.

We note that large values of Seebeck coefficient for the attractive model were earlier found in the DMFT~\cite{osolin15} but were not compared with the thermodynamic estimates and the importance of those results for the spin-thermoelectric response was not discussed.

The remainder of the paper is structured as follows. In Sec.~\ref{sec_model_method} we specify the model, the methods, and the notation. In Sec.~\ref{sec_results} we show our main results for the spin-Seebeck coefficient.  In Sec.~\ref{sec_dmft_transport} we describe the DMFT calculation of transport and in  Sec.~\ref{sec:attract} we exploit it in conjunction with the mapping to an attractive model to interpret our results. In Sec.~\ref{sec_diff} we investigate the influence of the spin-thermoelectric effect for the spin-diffusion. In Sec.~\ref{sec_conc} we give our conclusions. The Appendix discusses the behavior of spin-Seebeck coefficient for a phenomenological ansatz spectral function.

\section{Model and method}
\label{sec_model_method}
We study the square lattice Hubbard  model,
\begin{equation}
 H = -t\sum_{\langle i,j\rangle, s=\uparrow,\downarrow} c^\dagger_{i,s} c_{j,s}+U\sum_{i}
 n_{i,\uparrow}n_{i,\downarrow}
\label{eq_ham}
\end{equation}
with $t$ being the hopping between the nearest neighbors.  We take 
$\hbar=k_B=e=g\mu_B=1$. We likewise take lattice spacing $a=1$. We use $t$ as the energy unit.

We solve the Hamiltonian with FTLM~\cite{jaklic00,prelovsek13,kokalj13} on a $N=4\times 4$ cluster
and in the thermodynamic limit with the DMFT~\cite{georges96} (that is, we  solve the Hamiltonian Eq.~\eqref{eq_ham} in a local approximation).
The DMFT equations are solved using the 
numerical-renormalization group (NRG)~\cite{bulla08} in the
NRG-Ljubljana implementation~\cite{zitko09} as the impurity solver. 

\begin{figure}[ht!]
 \begin{center}
      \includegraphics[ width=0.99\columnwidth]{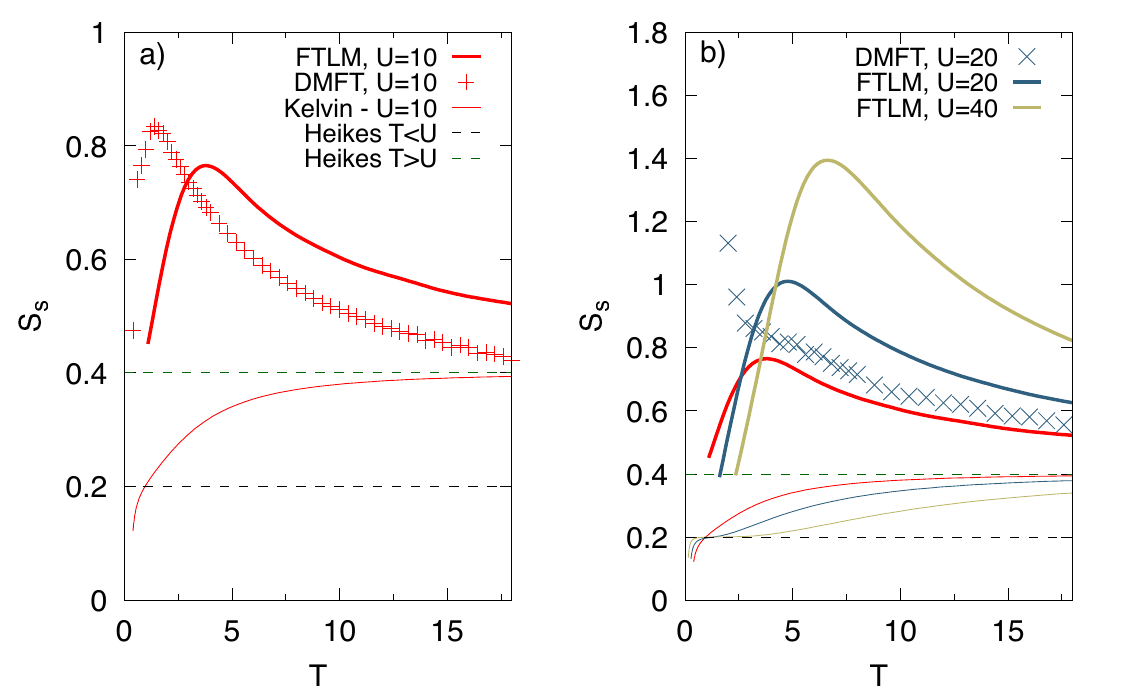} 
 \end{center}
 \caption{(a) The spin-Seebeck coefficient for $U=10$ at half filling for magnetization $m_z=0.05$, calculated using the FTLM (thick) and the DMFT
   (+). Kelvin estimates (thin) and the two Heikes' estimates are also
   shown (dashed). (b) The FTLM data for $U=10,20,40$ (thick) along the
   Kelvin estimates (thin). The DMFT data for $U=20$ are also shown (crosses). \label{fig1} }
\end{figure}

\begin{figure}[ht!]
 \begin{center}
   \includegraphics[ width=0.99\columnwidth]{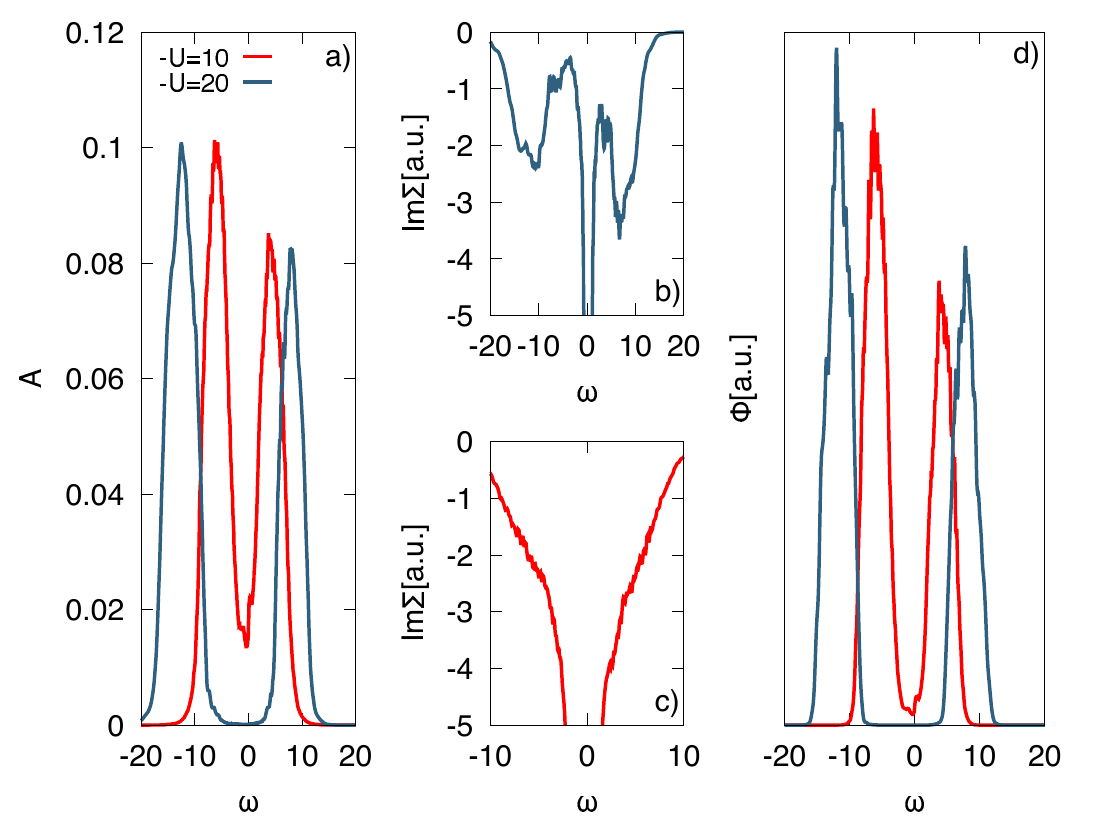} 
 \end{center}
 \caption{(a) The DMFT spectral function $A(\omega)$ for the doped attractive
   model $\delta=n-1=0.1$ (corresponding to $m_z=0.05$) for
   $-U=10,20$ at temperature $T=6$. (b,c) The corresponding imaginary part of the self energy,
   and (d) the DMFT transport function $\Phi(\omega)$. }
 \label{fig2}
\end{figure}

\section{Spin-Seebeck coefficient}
\label{sec_results}
Figure~\ref{fig1}(a) displays
$S_s$ as a function of temperature for $U=10$ evaluated with the FTLM
(full, thick) and the  DMFT (symbols). At highest temperatures, $S_s$
approaches the high-temperature Heikes value,
$2 \log (1+2 m_z)/(1-2m_z) \approx 8m_z$, which is the expected
behavior. The Kelvin estimate from $d B/dT|_{m_z}$ evaluated using
the FTLM (thin; DMFT gives similar results) agree with the Kubo evaluation
in this regime. On lowering the temperature, surprisingly, instead of
diminishing in amplitude as suggested by the Heikes value
corresponding to $T<U$,
$\log [(1+2m_z)/(1-2m_z)] \approx 4m_z$~\cite{chaikin76}, $S_s$
increases and reaches a maximum value well above the Heikes estimate,
and only drops consequently at lower $T$. This behavior with a
substantial increase of the spin-thermoelectric coefficient above the
high-temperature and thermodynamic estimates becomes even more
pronounced for larger $U$, as displayed in Fig.~\ref{fig1}(b).
The magnitude of the peak diminishes with decreasing magnetization but
increases with increasing $U$. It is, as we show below, proportional
to $m_z U/T$.

Throughout the considered regime, the temperatures are high ($T>t$)
and one cannot attribute the deviations from the thermodynamic
estimates to the occurrence of coherent transport or to a proximity to a magnetically
ordered regime. The observed deviations are in stark contrast to the
behavior of the charge thermopower that at temperatures $T>t$ does
follow the thermodynamic estimates~\cite{deng13}.

\section{DMFT description of transport}
\label{sec_dmft_transport}
In order to understand this behavior it is convenient to
discuss the transport properties within the DMFT approach. The DMFT
expresses the transport coefficients in terms of the transport
function~\cite{palsson98}
\begin{equation}
  \Phi_\sigma(\omega) = \sum_k v_k^2 A_{k\sigma}^2(\omega)
  \end{equation}
  where $v_k$ is the band velocity: $v_k=d \epsilon/dk_x$ with
  $\epsilon_k$ the band energy.  $A_{k\sigma}$ is the spectral
  function at momentum $k$ and spin $\sigma$.  The charge Seebeck $S_c$ and the
  spin Seebeck $S_s$ coefficients are, respectively,
\begin{equation}
  S_{(c,s)} =(1,2) \frac{\int[\Phi_\uparrow (\omega) \pm \Phi_\downarrow(\omega)] (-\omega/T) (-df/d\omega) d \omega}{\int [\Phi_\uparrow (\omega) + \Phi_\downarrow(\omega)] (-df/d\omega) d \omega} \label{eq:trans_int}.
\end{equation}

As seen in Fig.~\ref{fig1}(a), the DMFT results are similar, although not identical to the FTLM ones. To what extent the differences
are technical (it is quite challenging to converge the DMFT
calculations in this regime as discussed in Ref.~\cite{osolin15}) or
physical, such as emanating in non-local fluctuations and/or the
vertex corrections neglected in DMFT is a question that goes beyond
the scope of the present paper. It is likely that at low $T$ the
vertex corrections become more important as the DMFT calculation
gives an insulator with a spin gap whereas the actual behavior is that of a spin conductor described by a Heisenberg model.  For our purpose we will ignore these
differences between the two methods and exploit the more transparent
DMFT formulation of transport to interpret the FTLM results.

\section{Mapping to the attractive model} \label{sec:attract}
It is convenient to analyze the results in terms of the mapping of the
spin to the charge degrees of freedom for an attractive model, that is
$S_s (U) =2 S_c(-U)$, with the spin polarization
$2 m_z \rightarrow \delta$ becoming the
charge doping with factors of $2$ occurring due to the definition of spin.

 Figure~\ref{fig2}(a) presents the local spectral function $\sum_k A_k(\omega)$ of the doped
 attractive model at $T=6$, for two values of attraction
 $-U=10,20$. As discussed in earlier studies of the
 attractive model~\cite{keller01,capone02,kyung06,osolin15}, the spectral function consists of two peaks,
 which are as for the repulsive Hubbard model centered at $- \mu$ and
 $-\mu+U$, with $\mu\sim U/2$. The important distinction between the doped attractive and
 the doped repulsive model is in the behavior of the chemical potential
 with temperature that can at high-$T$ be most simply obtained from a
 grand-canonical treatment of the atomic problem. There, average
 electron occupancy can be evaluated from
\begin{equation}
  n = \frac{2 \exp(\beta \mu) +  2 \exp(-\beta (U- 2\mu) )}{1+2 \exp(\beta \mu) +   \exp(-\beta (U- 2\mu) )}
\end{equation}
In the attractive model terms that include $\exp(-\beta U)$  grow  at low temperatures and should be retained. 
Hence one obtains 
\begin{equation}
  \mu=U/2 + \delta\mu = U/2 + T \log((1+\delta)/(1-\delta))
  \end{equation}
  The fact that $\mu \sim U/2$ at low $T$ represents a crucial difference with respect to the repulsive case causes the spectral function to be gapped at  a nonvanishing doping. 
  Namely, the spectral function consists of two peaks displaced by $\delta \mu$ from $\pm U/2$ as shown on Fig~\ref{fig2}(a) for two
  values of 
  $U$.  Because at large $|U|$ the gap is well
  developed, the lower and upper Hubbard bands must have unequal
  spectral weight to yield a finite doping.  

Figure~\ref{fig2}(d) presents the transport function. One sees
that this exhibits the two Hubbard bands and is overall similar to
the density of states.  There is however an important difference:
Because $\Phi$ contains $A_k^2$, the weights of the upper and the
lower Hubbard band parts are affected by the amplitude of
scattering, as given by self-energy depicted in Figs.~\ref{fig2}(b) and \ref{fig2}(c) for $|U|=20,10$, respectively. When the spectral function is a sharply peaked function,  $A_k^2 \sim A_k /(2
\pi \Gamma_k)$ with $\Gamma_k=-\textrm{Im}\Sigma(\omega_k)$, and hence the transport function is modulated by the 
value of the self-energy at the peak frequency $\omega_k$. In passing we note that this finding can be used to connect the bubble expression to the Boltzmann
calculation, see Refs.~\cite{georges21,gourgout21} for a recent discussion.

In the Appendix~\ref{sec_phenom}
we take advantage of the simple two-peaked structure of the transport function and find an expression 
\begin{equation}
S_c= -\frac{U}{2T} \frac{\tilde{\phi}_- - \tilde{\phi}_+}{\tilde{\phi}_- + \tilde{\phi}_+} + \frac{\delta \mu}{T}= S_1 +\frac{\delta \mu}{T}
\label{eq:seeb_ph}
\end{equation}
where the first term grows as $U/T$ and is proportional to a coefficient that is expressed in terms  of the effective weights of the  positive (negative) frequency peaks of the transport function  $\tilde{\phi}_{\pm}$. This coefficient (that is found to be approximately given by doping, see Appendix~\ref{sec_phenom}) grows with the difference in the scattering between the holes and the electrons. When this difference does not vanish, the behavior of the Seebeck coefficient differs from that of the thermodynamic expectations given by the second term in Eq.~\eqref{eq:seeb_ph}.

Let us relate this discussion to the repulsive case. At the particle-hole symmetry, the spectral functions of the repulsive and attractive case for respectively  $s=\uparrow,\downarrow$ are related by 
$A_{k \uparrow,\downarrow} (\omega)|_{U>0} = A_{k \uparrow,\downarrow} (\pm \omega)|_{U<0} $. That is, taking advantage of the mapping, Fig.~\ref{fig2} depicts $s=\uparrow$ components of the spectral function, the self-energy and the transport function, and the $s=\downarrow$ components can be obtained by $\omega \rightarrow -\omega$.
The different scattering between the electrons and the holes in the attractive model thus relates to a different scattering between the spin majority and the spin minority carriers in the repulsive model, and this in turn leads to $S_s \sim m_z U/T$ behavior that explains the enhancement over the thermodynamic estimates seen in Fig.~\ref{fig1}. 


 \section{Diffusion matrix, diffusion eigenvalues, and relevance for experiment}
\label{sec_diff}
Predicted large values of spin-Seebeck
coefficient at large $T$ and the increase of the peak value with
increasing $U$ (or the corresponding behavior of the charge-Seebeck coefficient in the attractive case) could be tested in future cold-atom experiments.
In the
introduction we also raised a possibility that the existing measurements of
spin-diffusion would be affected by spin-thermoelectric effects, which could account for values of the spin-diffusion and the spin-conductivity that were found to be larger than theoretically expected.

Let us try to estimate the influence of thermoelectric effects on spin-conductivity using a
hand-waving argument. At
finite magnetization, gradients of magnetic field are accompanied by a gradient
of energy, and, assuming thermalization, a gradient of
temperature, $\nabla T \sim \nabla E/c = m_z\nabla B/c$, with $c$ as the
specific heat. Temperature gradients drive the spin current via
the spin-thermoelectric effect. Writing the spin current as $j_s = - L_{ss} \nabla B - L_{sq}
\nabla T/T$, one has $j_s = - L_{ss} (1 - S_s m_z/c) \nabla B$. At $T\approx 3t$ (relevant to experiment~\cite{nichols19}), the values
of specific heat are of the order $0.3 k_B$, hence the correction
of the estimated spin-conductivity due to spin-thermoelectricity for
magnetization $0.05$ where $S_s\approx0.8$  would be at the 15\% level.
Importantly, because $S_s$ and $m_z$ are of equal sign, this estimate anticipates the spin conductivity is actually reduced compared to the case where spin thermoelectricity is neglected.

\begin{figure}[t]
 \begin{center}
      \includegraphics[ width=0.99\columnwidth]{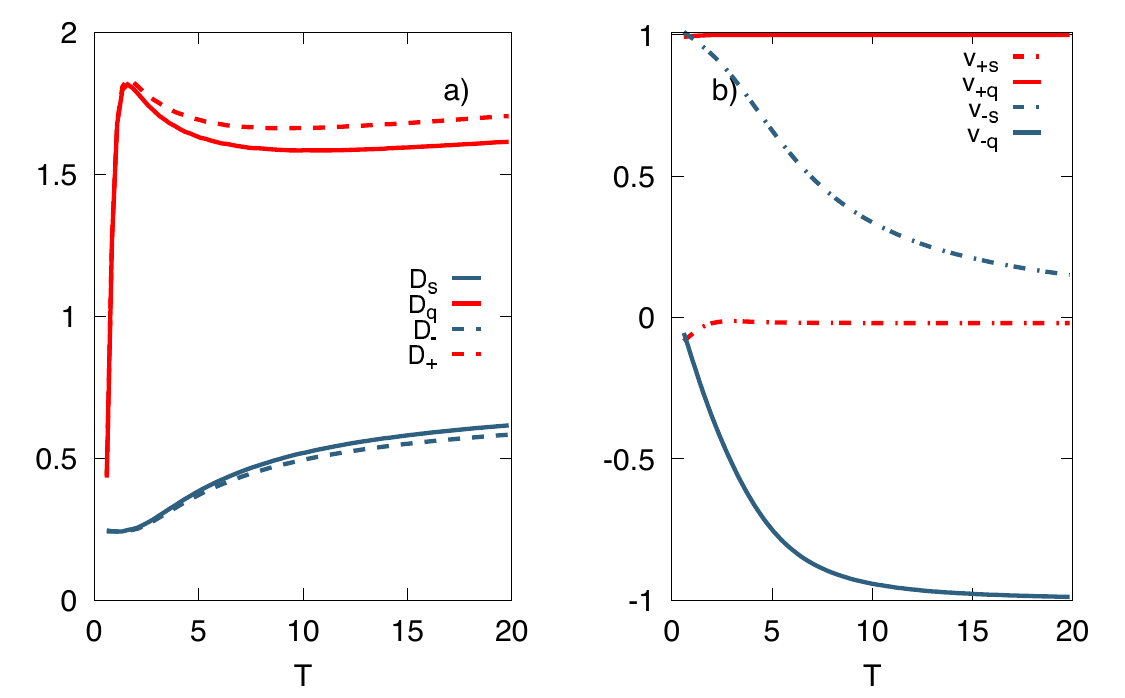} 
 \end{center}
 \caption{(a) Bare spin diffusion constant $D_s$, bare heat diffusion constant $D_q$
   and the two eigenvalues of diffusion matrix $\mathbf{D}$
   for $U=10$ and magnetization $m_z=0.05$. 
   (b) Spin and heat
   components of eigenmodes $\vec{v}_{\pm}$ at $m_z=0.05$.}
\label{fig4}
\end{figure}

In order to make this discussion more precise one must consider a generalization of the
Nernst-Einstein relation to a matrix formulation allowing for a mixed
response~\cite{hartnoll15},
where the off-diagonal entries involve
the mixed transport coefficient $L_{sq}$ and the thermoelectric susceptibility
$\xi=-\partial^2 f/\partial B\partial T$ where $f$ is the free energy density. 
The diffusion constant in
matrix form reads $\mathbf{D}=-\mathbf{LA}^{-1}$, where
$\lbrace {j}_q,{j}_s\rbrace=\mathbf{L}\lbrace \nabla T, \nabla B\rbrace$
defines the conductivity matrix $\mathbf{L}$ and the heat-magnetization susceptibility matrix $\mathbf{A}$ is
defined by $\lbrace T\nabla  s, \nabla m_z\rbrace=\mathbf{A}\lbrace \nabla T,\nabla B\rbrace$\cite{footnoteCharge}. 
The diffusion eigenmodes that are involved in general non vanishing spin and heat components are obtained by diagonalizing the matrix $\mathbf{D}$.
We denote  the diffusion eigenvalue whose mode contains a predominantly 
spin (heat) component by $D_-$ ($D_+$), respectively. These are shown in Fig.~\ref{fig4}(a) as a function of temperature for $U=10, m_z=0.05$ and are compared to the bare spin-diffusion constant
$D_s=\sigma_s/\chi_s$ and bare heat diffusion constant $D_q=\kappa/c$, where we use term ``bare'' to indicate that the spin-thermal mixing is neglected.

One sees that,
consistent with the hand-waving discussion above, the influence of the spin-thermoelectric
effects is only moderate, the diffusion eigenvalues are close to the values obtained when there is
no mixing. The behavior of  $D_\mp$ being smaller (larger) than the corresponding bare diffusion
 is that of level repulsion.

The relatively small admixing occurs because the geometric
mean of the off-diagonal elements
$\sqrt{D_{sq}D_{qs}}$ is significantly smaller ($\lesssim 10\%$) than $D_{qq}-D_{ss}$. When
any of the off-diagonal elements vanish, the spin-thermoelectric effect on
diffusion vanishes. We inspect more closely $D_{sq}$. It can be rewritten as $D_{sq}=-D_s\left(\frac{\xi +
S_s\chi_s}{c}\right)$, which expresses {\it spin-thermal diffusion}, i.e. spin-thermoelectric mixing between the spin and the thermal diffusion. 
Note that $S_s$ is multiplied by $\chi_s/c$, which becomes
large  at $T\rightarrow 0$. This hints at a rich behavior at low temperatures and should be explored in
future work. Using the Kelvin
formula $S_s^K= dB/d T|_{m_z}$, one can further rewrite:
$D_{sq}=-D_s\frac{\chi_s}{c}(S_s-S_s^K)$.
 Interestingly, when the spin Seebeck coefficient is given by the
 Kelvin estimate, $D_{sq}=0$, which leads to $D_-$ and $D_+$ equal $D_{ss}$ and $D_{qq}$.
In this case the spin modulation decays with pure spin diffusion constant
 $D_{ss}=\sigma_s/\chi_s$ and has an admixed thermal (heat) component.
An initial spin modulation therefore also induces heat currents.
On contrary, $D_+$ corresponds to pure heat diffusion decaying with the bare
 heat diffusion constant $D_{qq}=\kappa/c$.
 
In experiment, an initial gradient of magnetization is imposed and $D_{sq}$. Could larger values of the
experimentally inferred spin-diffusion constant occur because there is
a significant contribution of the heat eigenmode in the initial state that decays faster (see larger values of $D_q$ and $D_+$ in Fig.~\ref{fig4})?
The components of eigenvectors $\vec{v}_{\pm}$ are shown on Fig.~\ref{fig4}(b). At
$T=3.1$ we find the spin dominated eigenvector 
$\vec{v}_-=\lbrace -0.53 t, 0.84\rbrace$, i.e. it contains 
a significant component of heat current that only increases with temperature. (In this expression we reintroduced $t$ to indicate that the two components of $\vec{v}$ are in different units; because $t$ is a natural unit for the energy it is meaningful to compare the numerical values of the two components setting $t=1$).  
At the same temperature, the heat dominated 
$\vec{v}_+=\lbrace 0.999 t, -0.02\rbrace$
is mostly single component.
One can first assume that the initial state is given by a pure magnetization gradient. Expanding
this profile in terms of ${\vec{v}}_\pm$ we find only a small part of magnetization is
contained in $\vec{v}_+$. To be specific, at $T=3.1$,
$\vec{v}_+$ contains $\sim 2\%$ of the initial spin modulation (the relative weight of
$\vec{v}_+$ is sizable, but it carries only a small magnetization). Hence, the faster
decay of the $\vec{v}_+$ cannot importantly affect the evolution of the magnetization. 
What if the heat gradient component is also initially present?  Using the estimate $\delta
T\sim \frac{m_z}{\chi_s}\delta m_z$ (describing the situation
 where the magnetic field responsible for $\delta m_z$ is switched off and excess energy is
 instantly converted into heat), we find that $\vec{v}_+$
is more prominent in the initial state, but still accounts for 
$\sim 4\%$ of the initial magnetization modulation. 
In both cases, the majority of spin diffusion is therefore governed by $D_-\sim D_s$.

\section{Conclusions}
\label{sec_conc}
In summary, we calculated the spin-Seebeck
coefficient in the Hubbard model and discovered a rich behavior with
temperature. The spin-Seebeck coefficient significantly exceeds the
entropic estimate. This occurs due to the unequal scattering of spin
minority and spin majority carriers which gives rise to an $\propto
m_zU/T $ dependence that adds up to the Heikes' estimate.
 This is a striking demonstration of
the breakdown of the entropic interpretation of the thermopower in a
high-temperature regime where \textit{a priori} one would trust it the
most. Our predictions for a large spin-thermoelectric effect could be tested on
optical lattices. We also calculated the diffusion matrix eigenvalues
and estimated the influence of spin thermoelectricity in the existing
measurements of the spin-diffusion~\cite{nichols19}. We found this
influence to be moderate and insufficient to explain the discrepancy
between the experiment and the theory. Possible directions for future
research include simulating explicitly the time dependence in such
experiments and the study of possible nonlinear effects.  

\appendix

\section{Phenomenological analysis}
\label{sec_phenom}
 Here we take advantage of the known general shape of the transport
function to obtain an approximate simple expression for the Seebeck
coefficient.  The temperatures we consider in our simulations (and that pertain to the cold atom
experiments, which motivate our investigation) are large. In
Ref.~\cite{nichols19} the estimated entropy is 1.1, which pertains
to $T\approx0.3 |U|$ (in this Appendix we will phrase the discussion in terms of the attractive Hubbard model), comparable to the bandwidth.  Hence
the derivative of the Fermi function $df/d\omega$ does not change
substantially over each of the Hubbard bands and hence in the
transport integrals, Eq.~\eqref{eq:trans_int}, one can approximate the transport function by 
two $\delta-$peaks  as
\begin{equation}
\Phi(\omega) = \phi_- \delta(\omega - U/2+\delta \mu) + \phi_+ \delta(\omega + U/2+\delta \mu),
\end{equation}
with different weights $\phi_-$ and $\phi_+$ for the negative and positive
frequency peaks, respectively. Note that $U<0$ in this case. One can evaluate the Seebeck coefficient using this
ansatz transport function. Taking also into account that $\delta \mu
/T$ is small, one can expand the derivative of the Fermi function, that
is $-df/d\omega (\pm U/2 - \delta \mu)= -df/d\omega (U/2) (1 \pm  t_0
\delta \mu /T)$ where we define $t_0= \tanh(U/4T)$.  Introducing
  effective weights (modified from $\phi$ due to $df/d\omega$)
  $\tilde{\phi}_\pm = \phi_\pm (1 \mp t_0 \delta \mu/T )$,
 one obtains a
  simple expression for the Seebeck coefficient 
\begin{equation}
S_c= -\frac{U}{2T} \frac{\tilde{\phi}_- - \tilde{\phi}_+}{\tilde{\phi}_- + \tilde{\phi}_+} + \frac{\delta \mu}{T}= S_1 +\frac{\delta \mu}{T}
\label{eq:seeb_ph2}
\end{equation}

The high-$T$ Seebeck coefficient for the attractive model thus has a
Heikes' term (second term of this expression, predicted by Chaikin and
Beni~\cite{chaikin76}) but crucially also the first term $S_1$, that
is proportional to $U/2T$ and the difference between the effective
weights of the peaks of the transport function. Whenever this
difference (that, as we discuss next, is due to a different scattering 
of electrons and holes) does not vanish, the Seebeck
coefficient {\it cannot} be interpreted in terms of the entropic
considerations alone.  This explains large values of spin-Seebeck
coefficient seen in numerical results of Fig.~\ref{fig1}.

\begin{figure}[t!]
 \begin{center}
      \includegraphics[ width=0.99\columnwidth]{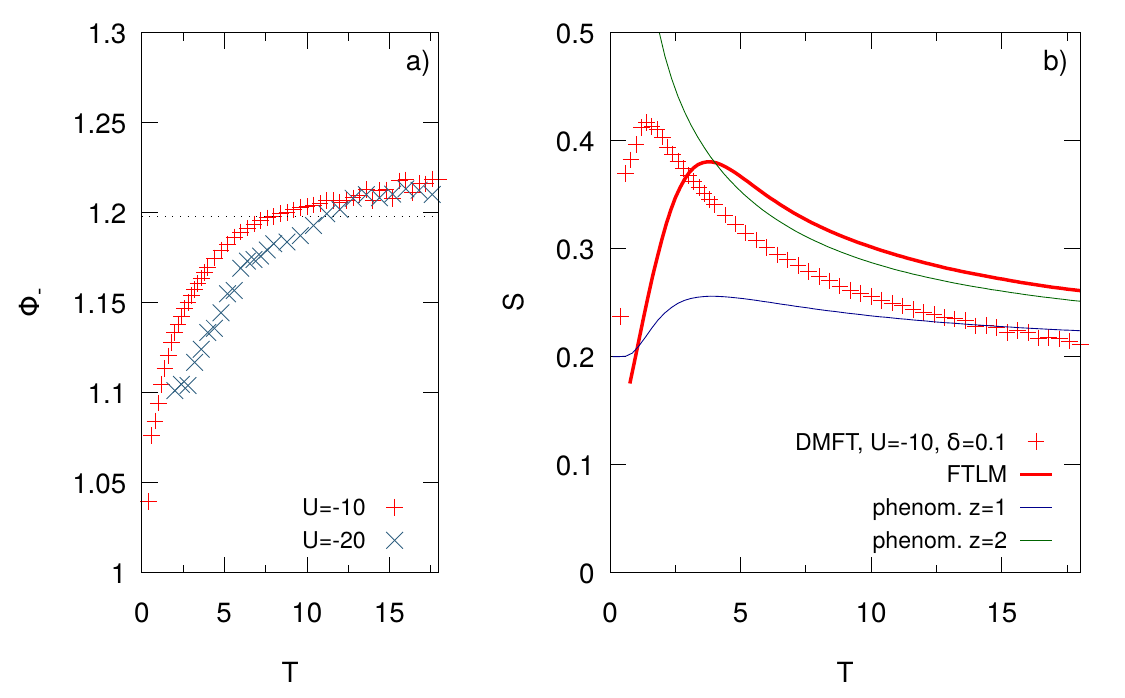} 
 \end{center}
 \caption{ (a) Weight of the negative frequency peak of transport function $\phi_-$ for attractive Hubbard model for $\delta=0.1$ and two
   values of $U$. The data are normalized such that total weight $\phi_-+\phi_+=2$. The value corresponding to the phenomenological estimate $\phi_-=(1+\delta)^z$ 
   with $z=2$ is also indicated (dashed). 
  (b) Seebeck coefficient for $U=-10$  compared with estimates
   based on the assumption of the transport function weights being
   $\phi_{\pm} = (1 \mp \delta)^z$ for $z=1,2$.
 }
\label{fig3}
\end{figure}

Is scattering really important? Would not the effective weights differ
already due to the different weights of the corresponding Hubbard band
weights in the density of states? If this were the case, one would
have $\phi_\pm \propto (1 \mp \delta)$ 
(simply from the considerations of
occupancy). Incidentally, the influence of $\delta \mu$ just cancels at
small $T$. Namely, for small doping $\delta \mu /T \approx \delta$. As
$T/|U|$ becomes small, $t_0 \rightarrow -1$. Hence one has
$\tilde{\phi}_\pm \approx \phi_\pm (1  \pm \delta) $, and $\tilde{\phi}_+
= \tilde{\phi}_-$. In this limit $S_1$ would vanish. One needs the
to take the scattering into account to understand the occurrence
of deviation from entropic estimates.

We plot the weight $\phi_-$, obtained from the integral of the transport
function over negative frequencies normalized such that the total
integral is 2 on Fig.~\ref{fig3}(a). One sees that in most of the
studied temperature range $\phi_{-}$ is close to the value expected from
the dependence $(1+\delta)^2$. Only at smaller temperatures the weight
$\phi_-$ decreases and actually approaches a smaller value $(1+\delta)$.

In Fig.~\ref{fig3}(b) we compare numerical results for $S_c$ with the
result of Eq.~\eqref{eq:seeb_ph2} where we set  $\delta \mu/T$ to the high-temperature Heikes value we approximate the transport function weights with 
$\phi_\pm \propto
(1 \mp \delta)^z$ with $z=1$ (blue) and with $z=2$ (green). At small 
temperatures these lead to a behavior  $S_1=-U/2T  (z-1) \delta$ (where
 corrections of order $\delta^2$ and higher are ignored).
For $z=1$, which corresponds
to taking into account just the different number of carriers, $S_1$ vanishes and the strong increase of $S_c$ seen in numerical simulations is not reproduced.  One needs to take into account also different scattering of carriers (as in this approximation 
embodied for $z=2$), too. 


\section*{Acknowledgements}
We acknowledge helpful discussions with Rok \v{Z}itko and Antoine Georges.
This work was supported by the Slovenian Research Agency (ARRS) under 
Program No. P1-0044 and Projects No. J1-2458, No. N1-0088, and No. J1-2455-1.


\begin{thebibliography}{34}
\bibitem [{foo()}]{footnoteCharge}%
  \BibitemOpen
  \href@noop {} {}\bibinfo {note} {Notice that we are considering an undoped,
  particle-hole symmetric case, hence the charge thermoelectric effect
  vanishes, thus we consider the Nernst-Einstein equation in terms of $2\times
  2$ matrices. Away from particle-hole symmetry charge transport effects would
  need to be considered as well and diffusion would be given in terms of
  $3\times 3$ matrices}\BibitemShut {NoStop}%
\end{thebibliography}

%


\end{document}